\documentclass[5p,twocolumn,times]{elsarticle}

\usepackage{lipsum}
\usepackage{lineno,hyperref}
\modulolinenumbers[5]
\usepackage[pdftex]{color}
\usepackage[font=footnotesize,labelfont=bf]{caption}
\usepackage[font=footnotesize,labelfont=bf]{subcaption}
\journal{Journal of \LaTeX\ Templates}

\usepackage{amssymb}

\biboptions{compress}

\usepackage[figuresright]{rotating}

\begin{document}

\begin{frontmatter}

\title{Ambush strategy enhances organisms' performance in rock-paper-scissors games}

\address[1]{School of Science and Technology, Federal University of Rio Grande do Norte, 
59072-970, P.O. Box 1524, Natal, RN, Brazil}

\address[2]{Edmond and Lily Safra International Institute of Neuroscience, Santos Dumont Institute,
Av Santos Dumont 1560, 59280-000, Macaiba, RN, Brazil}

\address[3]{Academy for Data Science and Artificial Intelligence, Zuyd University of Applied Sciences, Nieuw Eyckholt 300, 6419 DJ, Heerlen, The Netherlands}

\address[4]{Institute for Biodiversity and Ecosystem Dynamics, University of Amsterdam, Science Park 904, 1098 XH, Amsterdam, The Netherlands}

\author[1]{R. Barbalho} 
\author[1]{S. Rodrigues}  
\author[2]{M. Tenorio}  
\author[3,4]{ J. Menezes}

\begin{abstract}
We study a five-species cyclic system wherein individuals of one species strategically adapt their movements to enhance their performance in the spatial rock-paper-scissors game. 
Environmental cues enable the awareness of the presence of organisms targeted for elimination in the cyclic game.
If the local density of target organisms is sufficiently high, individuals move towards concentrated areas for direct attack; otherwise, they employ an ambush tactic, maximising the chances of success by targeting regions likely to be dominated by opponents. Running stochastic simulations, we discover that the ambush strategy enhances the likelihood of individual success compared to direct attacks alone, leading to uneven spatial patterns characterised by spiral waves. We compute the autocorrelation function and measure how the ambush tactic unbalances the organisms' spatial organisation by calculating the characteristic length scale of typical spatial domains of each species. We demonstrate that the threshold for local species density influences the ambush strategy's effectiveness, while the neighbourhood perception range significantly impacts decision-making accuracy. 
The outcomes show that long-range perception improves performance by over 60\%, although there is potential interference in decision-making under high attack triggers. 
Understanding how organisms' adaptation their environment enhances their performance may be helpful not only for ecologists, but also for data scientists, aiming to improve
artificial intelligence systems.
\end{abstract}

\end{frontmatter}

\section{Introduction}
\label{sec1}
Spatial interactions are crucial for species coexistence, as has been discovered in experiments with \textit{Escherichia coli} \cite{bacteria}. Researchers have found that three bacterial strains sustain survival through cyclic dominance, 
following the rules of the popular rock-paper-scissors game: scissors cut paper, paper wraps rock, rock crushes scissors \cite{mobilia2,Avelino-PRE-86-036112,uneven,sara}. However, maintaining biodiversity requires localised selection interactions, leading to distinct spatial domains \cite{Coli}. A similar phenomenon has been observed in Californian coral reef invertebrates and lizards in the inner Coast Range of California \cite{coral,lizards}. In these systems, organisms with limited mobility engage in local interactions, promoting coexistence and biodiversity. Conversely, increased mobility leads to a more uniform distribution of species, resulting in biodiversity loss \cite{Allelopathy}.

Plenty of evidence shows that in many biological systems, organisms adapt their actions based on environmental cues \cite{ecology,Causes}. Such behavioural strategies are crucial to organisms' survival, species persistence, and ecosystem stability \cite{MovementProfitable,Nature-bio}. 
For instance, animals exhibit adaptive movement patterns as a survival strategy under challenging situations or to locate resources and suitable habitats for reproduction. Many organisms can perceive and respond to environmental cues, adjusting their locomotion accordingly. The comprehension of behavioural mobility strategies has contributed to the development of advanced tools by engineers to improve robotic systems that mimic animal behaviour 
\cite{foraging,butterfly,BUCHHOLZ2007401,adaptive1,adaptive2,Dispersal,BENHAMOU1989375,coping}.

A common tactic among animals is the ambush behaviour, which aims to maximise the probability of an organism encountering and successfully eliminating prey or competitors \cite{ambush0,ambush00}.
For example, many marine zooplankton organisms, especially copepods, adopt an ambush feeding strategy known as "ambush", characterised by rapid and surprising attacks \cite{ambush2}. These attacks are enabled by precise manoeuvres during swift jumps, executed with high velocity and short duration, preventing the prey from escaping.

In Ref. \cite{ambush1}, the authors have discovered that organisms may drive their ambush tactic by rationalising prey features, like size, movement, and temperature. More giant vipers prefer larger prey, while prey movement and temperature influence striking behaviour, irrespective of whether the prey is alive or dead.
Researchers also have studied the relationship between ambush strategy and haematocrit (Hct), which measures the volume percentage of red blood cells in whole blood \cite{ambush3}. They discovered that snakes using ambush tactics have lower Hct levels than active foragers across different habitats, indicating that lower Hct levels may confer advantages in reducing maintenance and locomotory costs.

In the context of the rock-paper-scissors model, 
behavioural movement strategies play a vital role in species persistence and biodiversity maintenance \cite{tenorio1,tenorio2}. It has been shown that
defensive strategic movement, which allows organisms to escape being eliminated by enemies (Safeguard strategy \cite{Moura}) or being infected by a pathogen during epidemic outbreaks (Social Distancing \cite{combination,adaptivej,adaptivejj}) are efficient in protecting individuals, benefiting the species which predominates in the spatial game \cite{Moura,adaptivej,adaptivejj}.

In this work, we investigate the generalised rock-paper-scissors models wherein organisms from one species employ a locally adaptive Ambush strategy. This behavioural tactic starts by assessing the local abundance of individuals they dominate in the cyclic game by scanning the surrounding area. Subsequently, individuals perform a trade-off between two components of the Ambush strategy: the Attack and Anticipation movement tactics. This means that the Attack strategy is chosen only if the local density of target organisms meets or exceeds a predetermined threshold, ensuring successful attacks. Otherwise, individuals move towards areas where the target individuals are expected to be abundant shortly, anticipating their arrival to launch an attack.

We conduct our investigation by running stochastic simulations based on the May-Leonard implementation of the rock-paper-scissors model. In this framework, organisms engage in local interactions, and the population size is not conserved \cite{Menezes_2023,tanimoto2,Szolnoki-JRSI-11-0735, Anti1,anti2,MENEZES2022101606,PhysRevE.97.032415,Bazeia_2017,PhysRevE.99.052310}. We aim to address the questions: i) What is the influence of the Ambush strategy on spatial patterns? ii) How does the Ambush strategy affect the sizes of typical spatial domains occupied by individuals? iii) How does the performance of organisms in the rock-paper-scissors game vary based on the threshold used to determine the adequacy of local density for initiating the Attack strategy? iv) How does the organisms' perception radius impact the trade-off between the Anticipation and Attack components of the Ambush strategy? v) How does the optimal threshold leading to the most significant relative increase in organisms' game performance, change with alterations in the perception radius?	

The outline of this paper is as follows: the model and the methodology are introduced in Sec.~\ref{sec2}, where the simulations are explained in detail. 
The unevenness provoked by the Ambush strategy in the spatial patterns is observed in Sec.~\ref{sec3}, while the autocorrelation function and domains' characteristic length scales are computed in Sec.~\ref{sec4}. 
In Sec.~\ref{sec5}, we calculate the organisms' performance in the rock-paper-scissors game for every species. We study the role of the perception radius in the relative variation of the performance in the game in Sec.~\ref{sec6}. Finally, our discussions and conclusions appear in Sec.~\ref{sec7}.

\begin{figure}
\centering	
\includegraphics[width=40mm]{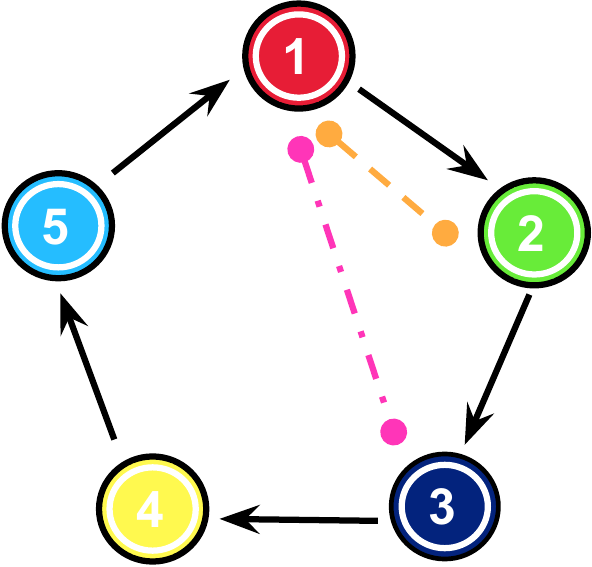}
\caption{Illustration of the rock-paper-scissors model with $5$ species. Selection interactions are represented by arrows indicating the dominance of organisms of species $i$ over individuals of species $i+1$. The two components of the Ambush strategy are illustrated by the orange dashed (Attack) and pink dotted-dashed lines (Anticipation), where organisms of species $1$ move towards the direction with a higher density of species $2$ and $3$, respectively.}
\label{fig1}
\end{figure}
\begin{figure*}
	\centering
    \begin{subfigure}{.19\textwidth}
        \centering
        \includegraphics[width=34mm]{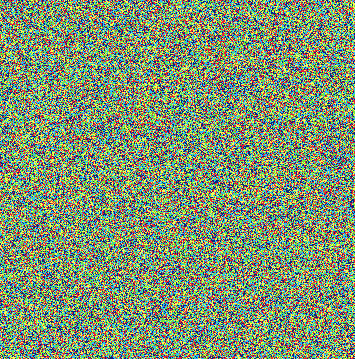}
        \caption{}\label{fig2a}
    \end{subfigure} %
   \begin{subfigure}{.19\textwidth}
        \centering
        \includegraphics[width=34mm]{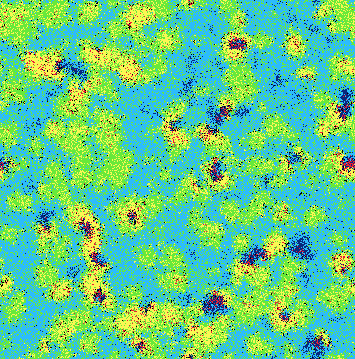}
        \caption{}\label{fig2b}
    \end{subfigure} 
            \begin{subfigure}{.19\textwidth}
        \centering
        \includegraphics[width=34mm]{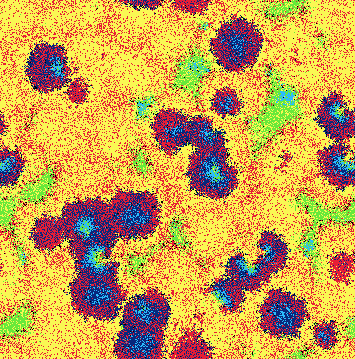}
        \caption{}\label{fig2c}
    \end{subfigure} 
           \begin{subfigure}{.19\textwidth}
        \centering
        \includegraphics[width=34mm]{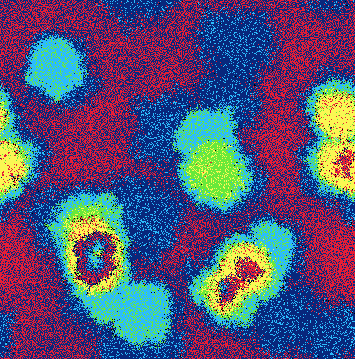}
        \caption{}\label{fig2d}
    \end{subfigure} 
   \begin{subfigure}{.19\textwidth}
        \centering
        \includegraphics[width=34mm]{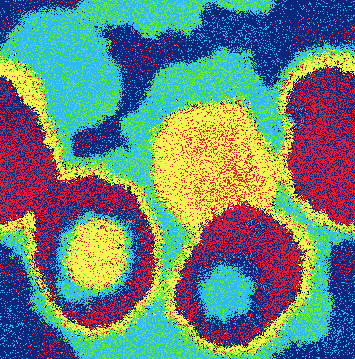}
        \caption{}\label{fig2e}
            \end{subfigure}\\
                \begin{subfigure}{.19\textwidth}
        \centering
        \includegraphics[width=34mm]{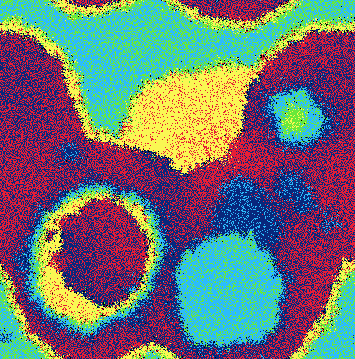}
        \caption{}\label{fig2f}
    \end{subfigure} %
   \begin{subfigure}{.19\textwidth}
        \centering
        \includegraphics[width=34mm]{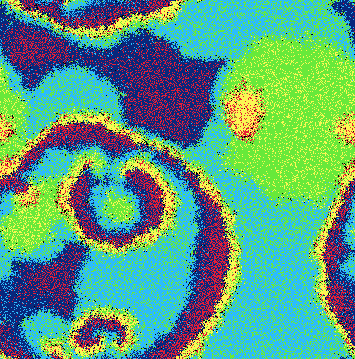}
        \caption{}\label{fig2g}
    \end{subfigure} 
            \begin{subfigure}{.19\textwidth}
        \centering
        \includegraphics[width=34mm]{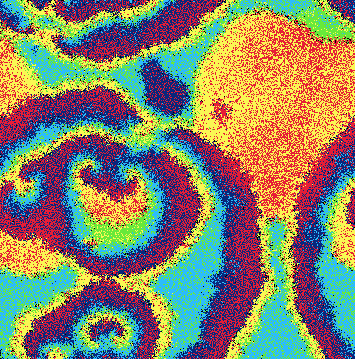}
        \caption{}\label{fig2h}
    \end{subfigure} 
           \begin{subfigure}{.19\textwidth}
        \centering
        \includegraphics[width=34mm]{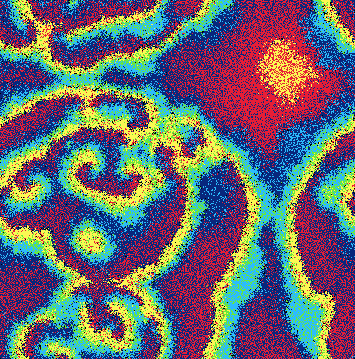}
        \caption{}\label{fig2i}
    \end{subfigure} 
   \begin{subfigure}{.19\textwidth}
        \centering
        \includegraphics[width=34mm]{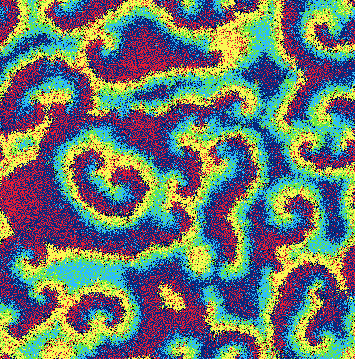}
        \caption{}\label{fig2j}
            \end{subfigure}
 \caption{Images captured during a simulation of the rock-paper-scissors model commencing from the random initial conditions in Fig.~\ref{fig2a}, with individuals belonging to species $1$ employ the Ambush strategy. The lattice has $500^2$ grid sites, the timespan of $5000$ generations, $R=5$ and $\beta=0.1$.
The organisms' spatial organisation at $t=140$, $t=280$, $t=560$, $t=880$, 
$t=1140$, $t=1760$, $t=260$, $t=2360$, and $t=4760$,
 generations are showed in Figs.~\ref{fig2b}, ~\ref{fig2c}, ~\ref{fig2d}, ~\ref{fig2e}, ~\ref{fig2f}, ~\ref{fig2g}, ~\ref{fig2h}, ~\ref{fig2i},and ~\ref{fig2j}. The colours follow the scheme in Fig~\ref{fig1}; empty spaces appear as white dots. The dynamics of the organisms' spatial distribution is shown in the video https://youtu.be/FK1u3ujm1LM.}
  \label{fig2}
\end{figure*}
\section{Model and methods}
\label{sec2}
		
Our model constitutes a cyclic relationship of five species whose competitive dynamics are governed by the rules illustrated in Fig.\ref{fig1}. Accordingly, individuals of species $i$ exhibit a selection interaction on individuals of $i+1$, where $i$ ranges from $1$ to $5$ - the cyclic identification is expressed as $i = i + 5\,\alpha$, where $\alpha$ denotes an integer. 
This formulation represents a direct extension of spatial rock-paper-scissors models to five species \cite{ramylla}.

In the course of our investigation, individuals of species $1$ execute a locally adaptive Ambush strategy with the overarching goal of maximising their performance in the cyclic game. This behavioural tactic is categorised into three sequential steps:
\begin{enumerate}
\item
Every organism of species $1$ surveys its local surroundings to assess the abundance of individuals of species $2$.
\item
Suppose the local density of organisms of species $2$ is greater than or equal to a predetermined threshold. In that case, the individual walks towards the direction with the highest density of individuals of species $2$. This behavioural movement is referred to as the Attack strategy \cite{Moura}.
\item
Otherwise, if the local density of individuals of species $2$ is inferior to the threshold, the organism activates the Anticipation strategy, characterised by a trajectory towards the direction with the highest abundance of species $3$ \cite{Moura}. The goal is to migrate to regions where individuals of species $2$ are expected to dominate soon, leveraging their advantage in the cyclic game. Consequently, individuals of species $1$ prepare to launch an attack, swiftly eliminating them upon their arrival.
\end{enumerate}

In our model, the Ambush strategy is locally adaptive, as each individual of species $1$ autonomously determines the most suitable movement strategy in real-time. Organisms from other species do not exhibit any specific behavioural movement strategy; instead, they move randomly, aligning with the conventional approach found in typical studies of spatial rock-paper-scissors models \cite{mobilia2}.

Our numerical simulations run on square lattices with linear size $N$ and periodic boundary conditions. 
The total number of grid points is $N^2$, which is the maximum quantity of organisms permissible in the system: each grid point accommodates at most one individual. 

We follow the May-Leonard model to implement the simulations, wherein the total number of individuals is not conserved \cite{leonard}. Initially,
an equal number of organisms of every species is
randomly distributed across the lattice. Denoting the total number of individuals as $I_i$, we express the initial count as $I_i = \mathcal{N}/5$ - where $i=1,2,3,4,5$ ensures the absence of empty spaces in the initial conditions.

Using the von Neumann neighbourhood, we consider that an organism can interact with one of its four immediate neighbours. The possible actions follow the rules:
a) Selection: $i\ j \to i\ \otimes$, where $j = i+1$, and $\otimes$ means an empty space;
b) Reproduction: $i\ \otimes \to i\ i$;
c) Mobility: $i\ \odot \to \odot\ i$, with $\odot$ representing either an individual of any species or an empty site.

The incidence of a particular interaction in our stochastic simulations is determined by a set of probabilities: $s$ (selection), $r$ (reproduction), and $m$ (mobility). These parameters remain uniform for all organisms of every species.
The implementation unfolds in three steps: i) a random choice of an active organism from the individuals in the grid; ii) a random choice of an interaction based on the set of probabilities; iii) 
a random choice of one of the four nearest neighbours to suffer the action.
The only exception is the Ambush strategy performed by organisms of species $1$, where the individual's local environment determines the direction of movement.

The algorithm implements the Ambush strategy in two steps: i) the choice of what movement tactic to execute - Attack or Anticipation; ii) the identification of the best direction to execute the chosen movement tactic.

The first step is implemented as follows:
\begin{enumerate}
\item Definition of the perception radius $R$, measured in lattice spacing, as the maximum distance an individual can scan environmental cues.
\item 
Implementation of a disc of radius $R$, centred at the active individual, outlining the total grid points the organism can analyse.
\item 
Definition of a real parameter $\beta$, where $0,\leq,\beta,\leq,1$, as the Attack trigger parameter. This threshold is considered by organisms of species $1$ as the minimum local density of individuals of species $2$ required to perform the Attack tactic.
\item
Computation of the density of individuals of species $2$ within the disc of radius $R$ surrounding the active individual. 
\item
If the local density is greater or equal to the threshold $\beta$, the organism opts for the Attack strategy; otherwise, the Anticipation tactic is chosen.
\end{enumerate}

The second step followed by the algorithm is described as:
\begin{enumerate}
\item Division of the organism's perception area into four circular sectors in the directions of the nearest neighbours. 
\item
Calculation of the number of target organisms within each circular sector - individuals on the borders are considered to be in both circular sectors. Here, the target organisms depend on the movement tactic to be implemented: individuals of species $2$ or $3$, in case of Attack or Anticipation strategies, respectively.
\item
Determination of the direction to walk based on the chosen strategy: the direction with the more significant number of the target organisms. In case of equal attractiveness among multiple directions, a drawn is conducted. 
\end{enumerate}

The results presented in this work originated from simulations conducted on square lattices of $500^2$ sites running for $5000$ generations. 
The interaction probabilities are assumed as follows: $s = r = m =1/3$. However, we have ascertained the generalisability of our conclusions across different sets of interaction probabilities by repeating simulations with alternative parameters.
\section{Spatial Patterns}
\label{sec3}
We started our investigation with a single simulation to thoroughly investigate the spatial patterns emerging from the initial conditions in the rock-paper-scissors model (Fig.~\ref{fig1}). In this simulation, individuals of species $1$ execute the Ambush strategy with a perception radius of $R=5$ and $\beta=0.1$.
Figures \ref{fig2a}, \ref{fig2b}, \ref{fig2c}, \ref{fig2d}, \ref{fig2e}, \ref{fig2f}, \ref{fig2g}, \ref{fig2h}, \ref{fig2i}, and \ref{fig2j} illustrate the spatial organisation of organisms at: $t=0$,
$t=140$, $t=280$, $t=560$, $t=880$, $t=1140$, $t=1760$, $t=260$, $t=2360$, and $t=4760$, respectively.
 
Following the colour scheme in Fig.~\ref{fig1}, organisms of species $1$, $2$, $3$, $4$, and $5$ are represented by red, green, dark blue, yellow, and light blue, respectively, while black dots denote vacant spaces. The dynamics of the spatial organisation during the whole simulation are available in the accompanying video: https://youtu.be/FK1u3ujm1LM.

The simulation begins with random initial conditions (Fig.~\ref{fig2a}), leading to frequent selection interactions during the initial stages. Due to the unevenness introduced by the Ambush strategy of individuals in species $1$, there is an alternating dominance of territories among species in the initial transient stage of pattern formation. This spatial effect can be observed in Figures \ref{fig2b} to \ref{fig2f}, where groups of organisms of non-competing species create planar waves that spread in all directions, forming expanding domains that are subsequently replaced by species that dominate in the rock-paper-scissors game. Specifically, regions reached by wavefronts of species ${3,5}$ are later occupied by groups of individuals of species ${2,5}$, ${2,4}$, ${1,4}$, and ${1,3}$ - then, the cycle repeats.

As time progresses, one observes an increasing asymmetry between the average size of the typical spatial domains of different pairs of species. For example, in Fig.~\ref{fig2e}, the average width of the wavefront of spatial domains of species ${1,4}$ (red and yellow) is shorter than the wavefront of ${1,3}$ (red and dark blue). As a consequence of the collision of planar waves, spiral waves form and spread across the entire lattice, as observed in Figures \ref{fig2g} to \ref{fig2j}. The irregular spiral formation reflects the turbulence in the spatial segregation of species when individuals of species $1$ no longer move randomly but towards the best direction according to the Ambush strategy.


\begin{figure}
   \centering
  \includegraphics[width=85mm]{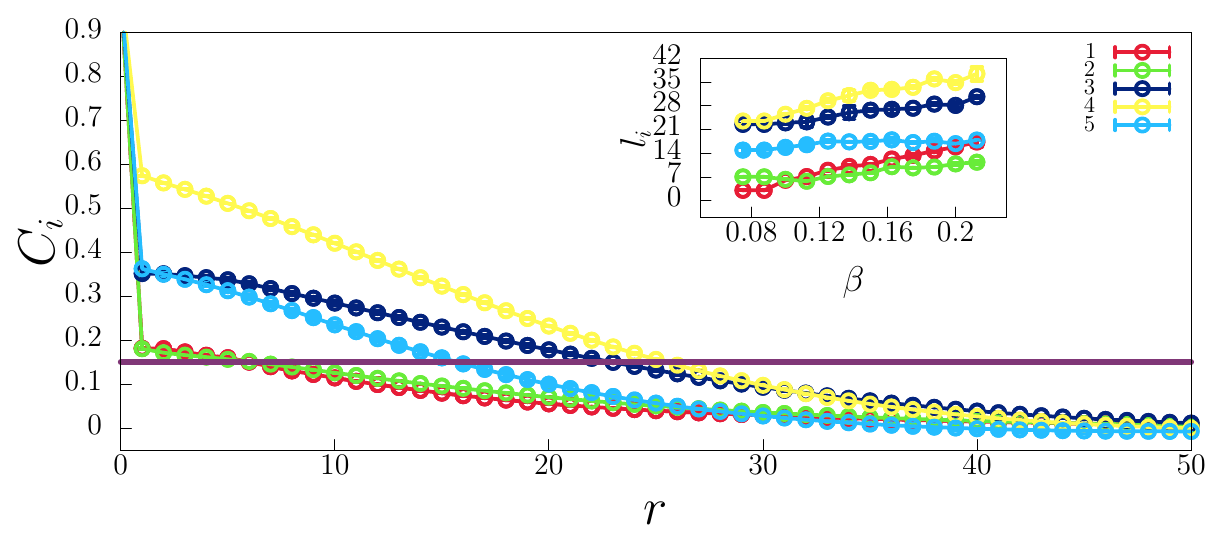}
  \caption{Autocorrelation function for every species in the rock-paper-scissors with Ambush strategy. We set the organisms' perception radius to $R=5$ and the Attack trigger to $\beta=0.1$. The outcomes were averaged from collections of $100$ simulations running in grid with $500^2$ sites until $5000$ generations. The colours follow the scheme in Fig.~\ref{fig1}; the error bars indicate the standard deviation. The inner panel shows the characteristic length scale for various values of $\beta$. The horizontal purple lines indicate the threshold assumed to calculate the characteristic length scale shown in the inner panel.}
 \label{fig5}
\end{figure}

\section{Spatial domains' characteristic length scale}
\label{sec4}

We now examine the impact of the Ambush strategy on the scale of typical spatial domains for individuals within the same species. To commence this inquiry, we determine the spatial autocorrelation function, denoted as $C_i(r)$, in terms of the radial coordinate $r$ for individuals of each species, where $i=1,2,3,4,5$ \cite{combination}.

To specify the position $\vec{r}$ in the lattice occupied by individuals of species $i$, we introduce the function $\phi_i(\vec{r})$. Utilizing the mean value $\langle\phi_i\rangle$ for the Fourier transform:
\begin{equation}
\varphi_i(\vec{\kappa}) = \mathcal{F}\,\{\phi_i(\vec{r})-\langle\phi_i\rangle\}, 
\end{equation}
and the spectral densities 
\begin{equation}
S_i(\vec{\kappa}) = \sum_{\kappa_x, \kappa_y}\,\varphi_i(\vec{\kappa}).
\end{equation}

Next, we perform the normalised inverse Fourier transform to obtain the autocorrelation function for species $i$ as
\begin{equation}
C_i(\vec{r}') = \frac{\mathcal{F}^{-1}\{S_i(\vec{k})\}}{C(0)},
\end{equation}
which can be written as a function of $r$ as
\begin{equation}
C_i(r') = \sum_{|\vec{r}'|=x+y} \frac{C_i(\vec{r}')}{min\left[2N-(x+y+1), (x+y+1)\right]}.
\end{equation}
The determination of the characteristic length scale for the spatial domains of species $i$, denoted as $l_i$ with $i=1,2,3,4,5$, involves setting the threshold $C_i(l_i)=0.15$.

Figure \ref{fig5} presents the results of a set of $100$ simulations for $R=5$ and $\beta=0.10$. The mean value of the autocorrelation function is depicted using the colours in the scheme of Fig.~\ref{fig1}, and the error bars show the standard deviation. The grey dashed line indicates the threshold used to compute the characteristic length scale $l_i$, defining the average size of the typical spatial domain occupied by each species. Consequently, the Ambush strategy performed by individuals of species $1$ creates an asymmetry in the spatial territory occupation, with individuals of species $4$ being more spatially correlated. The red and green dots also indicate that species $1$ and $2$ are the least spatially correlated.

To quantify the interference of the Attack trigger with spatial species correlation, we ran $100$ simulations with varying values of $\beta$ (ranging from 0.075 to 0.2125 in intervals of 0.0125). The individuals' perception radius was held constant at $R=5$; $l_i$, for $i=1,2,3,4,5$, was calculated using the threshold $C_i(l_i)=0.15$ - the purple horizontal line in Fig.~\ref{fig5}.
The outcomes of these simulations are depicted in the inner panel of Fig.~\ref{fig5}.

Examining the mean value of the characteristic length scale $l_i$ for $i=1,2,3,4,5$, we observe that organisms of species $4$ consistently form spatial domains with the most extended typical length scale, irrespective of the Attack trigger. The findings also reveal that as $\beta$ increases, the typical size of groups of individuals of the same species increases for every species. 

Furthermore, with low $\beta$, organisms of species $1$ occupy spatial patterns with the shortest characteristic length scale, and as the Attack trigger increases, $l_2$ becomes the shortest. This reversal happens because, as $\beta$ grows, organisms of species $1$ become more stringent in implementing the Attack part of the Ambush strategy, executing the Anticipation more frequently.
\begin{figure}
   \centering
  \includegraphics[width=85mm]{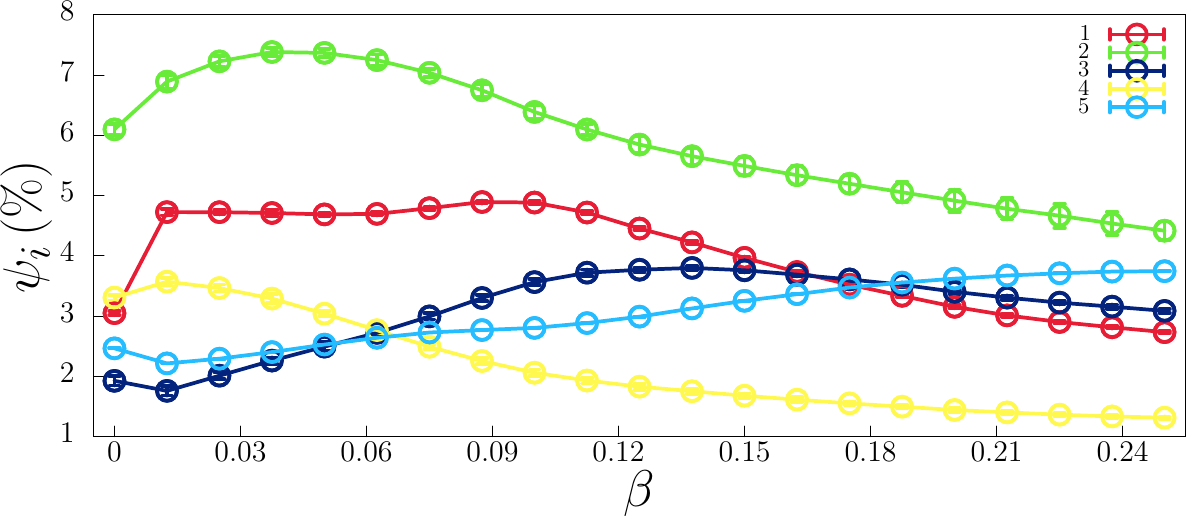}
  \caption{Organisms' performance in the spatial rock-paper-scissors with the Ambush strategy in terms of the Attack trigger. 
The colours indicate the species according to the scheme in Fig.~\ref{fig1}. The results were obtained by averaging the outcomes of $100$ simulations for each value of $\beta$, running in lattices with $500^2$ points for a timespan of $5000$ generations; the error bars depict the standard deviation. The perception radius was assumed to be $R=5$.}
 \label{pe}
\end{figure}

\section{The influence of the Attack trigger}
\label{sec5}

The Ambush strategy aims to maximise the individuals' efficiency in the cyclic spatial game. To quantify the efficacy of this tactic,
we assess the organisms' performance in the rock-paper-scissors game: $\psi_i$, defined as the probability of an individual belonging to species $i$ eliminating an organism from species $i+1$ within a one-generation interval.

For this purpose, we employ the following methodology:
\begin{enumerate}
\item
We tabulate the overall count of individuals of species $i$ at the onset of each generation;
\item
We ascertain the number of individuals from species $i+1$ that are killed by individuals of species $i$ during the generation;
\item
We compute $\psi_i$ as the ratio between the number of individuals from species $i+1$ eliminated during the generation and the initial total count of organisms belonging to species $i$.
\end{enumerate}

In this investigation, we constrain the parameter $\beta$ within the range $0 \leq \beta \leq 0.25$. This limitation for the interval of $\beta$ is made because, in scenarios where $\beta$ exceeds $0.25$, there is a significant reduction in the probability of organisms engaging the Attack strategy. This decline can be attributed to the low likelihood of an organism from species $1$ being surrounded by more than $25\%$ of conspecifics. Such circumstances arise due to the coexistence of organisms of five species alongside unoccupied spaces. Consequently, the efficacy of the Ambush strategy is compromised, leading organisms to predominantly opt for the Anticipation tactic over the Attack strategy in such contexts.

Figure \ref{pe} depicts sets of $100$ simulations for $R=5$ and $\beta=0.1$. The average performance in the game for each species is represented by the colour in Fig.\ref{fig1}, with error bars showing the standard deviation.
For $\beta \approx 0$, individuals of species $1$ always opt for the Attack strategy, as they require only a few individuals of species $2$ in their vicinity to perceive a local attractiveness for the attack. Consequently, the population of species $2$ diminishes, increasing the density of species $3$ \cite{Moura}. The low density of species $2$ individuals and the high proportion of organisms from species $3$ facilitate the selection activity: the ratio between the amount of individuals of species $3$ and $2$ is significantly high.
Because of this, the performance in the game $\psi_2$ experiences a significant increase, attributable to the execution of the Ambush strategy by organisms of species $1$ (as indicated by the green line in Fig.~\ref{pe}). This is the same rationale for $\psi_4>\psi_5$.

For high $\beta$, there is an augmented probability of organisms of species $1$ opting for the Anticipation strategy. This happens because of the reduced chances of the local density of species $2$ meeting the criteria required to initiate the Attack tactic. As a result, the performance in the game diminishes for individuals of species $1$, $2$, and $4$.

\section{The role of the perception radius}
\label{sec6}

Finally, our study seeks to clarify the impact of the perception radius, $R$, on organisms'  performance in the rock-paper-scissors game with the Ambush strategy.

. We now focus on individuals of species $1$, which are the only ones to perform the behavioural tactic.
For fulfilling this study, we introduce the relative variation in the organisms' performance, denoted as $\tilde{\phi}_1 (\beta)$. This parameter is intricately linked to the Attack trigger and is formally defined as
as
\begin{equation}
\tilde{\psi_1} = \frac{\psi_1 - \psi^0_1}{\psi^0_1},
\end{equation}
where $\psi^0_1$ represents the game performance corresponding to $\beta=0$.

We executed collections of $100$ simulations to attain this objective, systematically exploring a range of Attack triggers. The realisations were categorised into three groups: 
\begin{itemize}
\item For $R=3$, an organism can scan twenty-eight neighbouring grid sites. This configuration allows the assessment of local density for organisms of the target species at intervals of $1/28$. 
\item
For $R=4$, the perceptual range is expanded to observe the content of forty-eight surrounding grid points. Therefore, an individual of species $1$ can measure the local density of species $2$ with better accuracy, namely, in intervals of $1/48$.
\item
For $R=5$, the perceptual range encompasses eighty grid sites, thus improving even more the accuracy for evaluating the local density of target individuals in intervals of $1/80$.
\end{itemize}

\begin{figure}
   \centering
  \includegraphics[width=85mm]{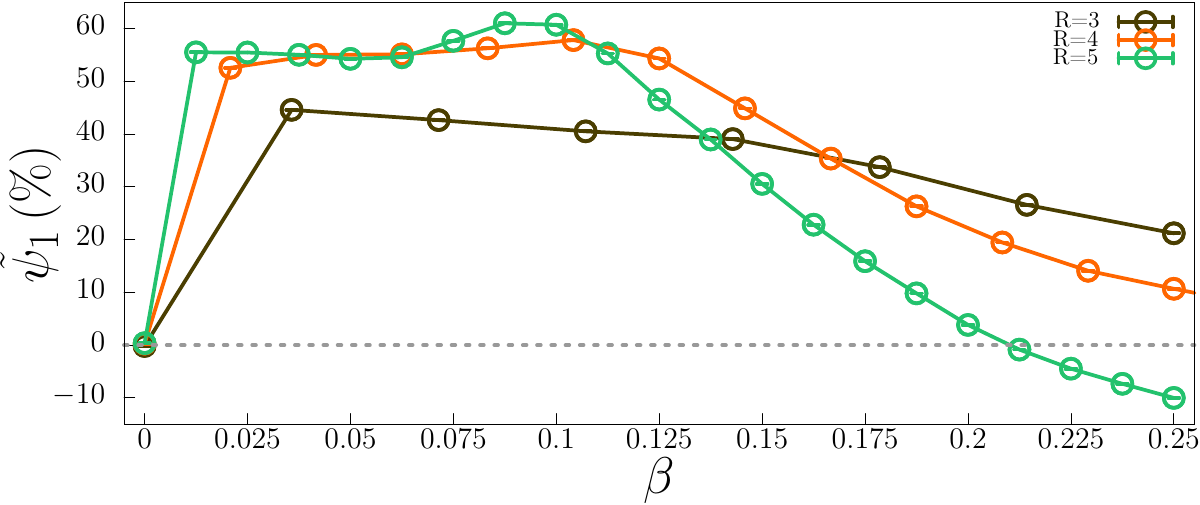}
  \caption{Relative variation in the organisms' performance in the rock-paper-scissors game for species $1$ as a function of the Attack trigger.
 The outcomes were averaged from sets of $100$ simulations starting from different initial conditions; the standard deviation is depicted by error bars. The grid size is $500^2$ and the timespan is $5000$ generations. The brown, orange and green lines show the results for $R=3$, $R=4$, and $R=5$, respectively; the dashed grey line indicates $\tilde{\psi}_1=0$}.
 \label{fig6}
\end{figure}

The relative variation in organisms' performance in the rock-paper-scissors game as a function of $\beta$ is depicted in Fig. \ref{fig6}. 
The outcomes were obtained by averaging the data from sets of $100$ simulations, with error bars representing the standard deviation. The brown line depicts the scenario where organisms of species $1$ can decide how to employ the Ambush strategy based on scanning and interpreting cues in the vicinity for $R=3$, while the orange and green lines represent scenarios with $R=4$ and $R=5$, respectively.

We initially observe that, regardless of the perception radius, the Ambush strategy enhances organisms' performance in the game for low $\beta$ compared to the scenario where individuals only employ the Attack tactic ($\beta=0$). Figure \ref{fig6} shows that, as the individuals' perception increases, the performance becomes more optimised. Furthermore, the results indicate the existence of an optimum Attack trigger, $\beta^{\star}$, determining the maximum $\tilde{\psi_1}$ which depends on the perception radius $R$.

We observe that for $\beta>\beta^{\star}$, a decline in the relative chance of organisms' performance in the game is observed.
For $R=3$, $\tilde{\psi_1} = 43.60 \%$ is achieved if the Attack trigger is $\beta^{\star}=1/28$. Thus, the performance reaches its maximum when an individual of species $1$ employs the Attack strategy, even if only one organism of species $2$ is detected. Conversely, suppose the organism is stricter, deciding to attack only in the presence of more than one individuals of species $2$. In that case, the performance decreases, but the Ambush strategy still brings positive outcomes, with $\tilde{\psi_1} >0$.

For $R=4$, there is an enhancement in the positive results obtained using the Ambush strategy. We found that, for $\beta^{\star}=5/48$, the organisms' performance growth is $\tilde{\psi_1} = 58.42\%$. Nonetheless, the best scenario is whether the individuals can perceive the neighbourhood in a radius of $R=5$: the performance in the rock-paper-scissors game enhances $\tilde{\psi_1} = 61.10 \%$, for $\beta^{\star}=8/80$. This implies that if $8$ organisms of species $2$ are detected among the $80$ positions scanned, it is time to cease anticipating and start attacking.

The dashed grey line indicates that, for $R=5$, the decline in $\tilde{\psi_1}$ is pronounced for $\beta>\beta^{\star}$. Furthermore, for $\beta>17/80$, the Ambush strategy becomes detrimental to organisms' performance in the game compared to employing the pure Attack strategy ($\beta=0$). This occurs because, when organisms scan long distances, there is a chance of being influenced by a distant collective of individuals from species $2$. Therefore, decisions may be based on inaccessible organisms, as each individual can move only one step within the lattice.

\section{Discussion and Conclusions}
\label{sec7}

Our study address a cyclic five-species system where organisms of species $i$ adapt their behaviour to enhance performance in a spatial rock-paper-scissors scenario. This involves environmental surveillance to identify potential targets (individuals of species $i+1$), thus employing the Ambush strategy. Namely, if the local density of target organisms suffice to ensure successful invasion, individuals move towards the direction in which the targets are mainly concentrated, executing the Attack strategy.
Conversely, when such density is lacking, individuals adopt the Anticipation strategy, positioning themselves strategically to intercept opponents likely to dominate the region.

We performed stochastic simulations and quantified the impact of the Ambush strategy of species $i$ in spatial pattern formation. Using the autocorrelation function, we show that the characteristic length scale of the typical species cluster of individuals of the same species is influenced by the threshold adopted by organisms of species $i$  to decide if the local density of species $i+1$ is high enough to perform the Attack strategy ($\beta$). The outcomes indicate that for low $\beta$, the groups of individuals of species $1$ are the smallest in the grid, growing as the Attack trigger increases.

Our findings revealed that implementing the Ambush strategies enhances the likelihood of individual success compared to executing exclusively the Attack tactic \cite{Moura}. Furthermore, we show that the Attack trigger and the organisms' perception radius significantly influence the performance in the spatial game. 
We found that as $R$ grows, organisms can more accurately find the optimal Attack trigger, which results in a higher relative increase in the performance in the rock-paper-scissors game. For $R=5$, the probability of an individual succeeding in the spatial game grows $61.10\%$ if assuming the optimal Attack trigger $\beta^{\star}=8/80$.
Finally, our findings show that, for excessive perception range, the organisms' performance may sharply drop because the decision to attack may be influenced by the perception of groups of distant individuals, which may not be realistic to access. 

Our stochastic simulations have uncovered a significant unevenness in cyclic games when a species gains strategic mobility. Rather than an equal distribution of natural resources and spatial territories among species, those employing behavioural ambush tactics may gain a notable advantage in the performance in the game. However, as elucidated in studies such as Ref. \cite{Moura}, which investigated strategic behavioural movement independently, this performance advantage can imperil species populations due to the inherent interactions governed by rock-paper-scissors cyclic rules.
To provide further precision, as indicated in the outcomes presented in Ref. \cite{Moura}, when considering a perception radius of $R=5$, a species population experiences approximately a $10\%$ decline if organisms exclusively move using the Attack strategy. This reduction becomes more pronounced, reaching roughly $36\%$, if organisms solely employ the Anticipation strategy. In our novel Ambush approach strategy, organisms operate based on their local realities, with some using the Attack tactic while others execute the Anticipation strategy. Consequently, species density may decline, on average, between $10\%$ and $36\%$, when compared with the standard model, where all organisms of every species move randomly.
Moreover, as organisms utilise environmental cues to determine movement directions based on the predetermined threshold $\nu$ assumed by organisms to assess the local density of target individuals, we conclude that the higher the attack triggers is, the lower the species population.  

The scenario becomes more intriguing concerning the density of empty space. Following the findings presented in Ref. \cite{Moura}, isolated execution of the Attack and Anticipation strategies yields contrasting consequences. While the Attack strategy leads to a slight increase of approximately $5\%$ in the density of empty spaces compared to the standard model, the Anticipation tactic results in a drop of approximately $50\%$. Consequently, in our Ambush strategy, the Attack trigger plays a pivotal role in determining the average number of empty spaces in the system: the higher $\nu$ is, the more available vacancies organisms encounter in the system, thereby enhancing the prospects of successful reproduction.

This highlights the potential for future investigations utilising our innovative Ambush strategy model. We aim to promptly explore and quantify the effects of local interactions on global population dynamics. Additionally, a critical ecological issue to be addressed is the impact of the Ambush strategy on the probability of coexistence in cyclic spatial games. In this regard, it is essential to delve into the underlying rationale guiding individuals to select a specific threshold for determining actions within the Ambush strategy. This exploration could be driven by optimisation goals such as maximising population density or minimising biodiversity loss.

Our model can also be extended to address other behavioural strategies involving, for example, organisms' reaction to the presence of enemies. 
By performing a trade-off between defence and attack movements	, an individual may opt to execute the Safeguard strategy \cite{Moura,Menezes_2024} instead of attacking if, after scanning and interpreting the environmental cues, the conclusion is that the local dangerousness surpasses the attractiveness for attacking.
In this case, organisms may benefit from increased survival time beyond better performance in the spatial rock-paper-scissors game.

\section*{Acknowledgments}
We thank CNPq/Brazil, ECT/UFRN, FAPERN/RN, IBED/UvA, ADSAI/Zuyd and Brightlands Smart Services Campus for financial and technical support.
\bibliographystyle{elsarticle-num}
\bibliography{ref}

\end{document}